\documentclass[twocolumn]{aastex631}
\usepackage[utf8]{inputenc}
\usepackage{CJKutf8}
\usepackage{url}
\usepackage{amsmath}
\usepackage{amssymb}
\usepackage{bm}
\usepackage[toc,page]{appendix}
\usepackage{listings}

\begin{document}

\title{Improving Harmonic Analysis using Multitapering:\\
Precise frequency estimation of stellar oscillations using the harmonic F-test}

\author[0000-0002-7626-506X]{Aarya A. Patil}
\correspondingauthor{Aarya A. Patil}
\email{patil@mpia.de}
\altaffiliation{LSST-DA Catalyst Fellow}
\affiliation{Max-Planck-Institut für Astronomie, Königstuhl 17, D-69117 Heidelberg, Germany}
\affiliation{David A. Dunlap Department of Astronomy \& Astrophysics, University of Toronto, 50 St George Street, Toronto ON M5S 3H4, Canada}
\affiliation{Dunlap Institute for Astronomy \& Astrophysics, University of Toronto, 50 St George Street, Toronto, ON M5S 3H4, Canada}

\author[0000-0003-3734-8177]{Gwendolyn M. Eadie}
\affiliation{David A. Dunlap Department of Astronomy \& Astrophysics, University of Toronto, 50 St George Street, Toronto ON M5S 3H4, Canada}
\affiliation{Department of Statistical Sciences, University of Toronto, 9th Floor, Ontario Power Building, 700 University Ave, Toronto, ON M5G 1Z5, Canada}

\author[0000-0003-2573-9832]{Joshua S. Speagle (\begin{CJK*}{UTF8}{gbsn}沈佳士\ignorespacesafterend\end{CJK*})}
\affiliation{Department of Statistical Sciences, University of Toronto, 9th Floor, Ontario Power Building, 700 University Ave, Toronto, ON M5G 1Z5, Canada}
\affiliation{David A. Dunlap Department of Astronomy \& Astrophysics, University of Toronto, 50 St George Street, Toronto ON M5S 3H4, Canada}
\affiliation{Dunlap Institute for Astronomy \& Astrophysics, University of Toronto, 50 St George Street, Toronto, ON M5S 3H4, Canada}
\affiliation{Data Sciences Institute, University of Toronto, 17th Floor, Ontario Power Building, 700 University Ave, Toronto, ON M5G 1Z5, Canada}

\author{David J. Thomson}
\altaffiliation{FRSC \& Emeritus Professor}
\affiliation{Department of Mathematics \& Statistics, Queen's University, Kingston, ON K7L 3N6, Canada}

\begin{abstract}
In Patil et. al 2024a, we developed a multitaper power spectrum estimation method, \texttt{mtNUFFT}, for analyzing time-series with quasi-regular spacing, and showed that it not only improves upon the statistical issues of the Lomb-Scargle periodogram, but also provides a factor of three speed up in some applications. In this paper, we combine \texttt{mtNUFFT} with the harmonic F-test to test the hypothesis that a strictly periodic signal or its harmonic (as opposed to e.g. a quasi-periodic signal) is present at a given frequency. This \texttt{mtNUFFT}/F-test combination shows that multitapering allows detection of periodic signals and precise estimation of their frequencies, thereby improving both power spectrum estimation and harmonic analysis. Using asteroseismic time-series data for the Kepler-91 red giant, we show that the F-test automatically picks up the harmonics of its transiting exoplanet as well as certain dipole ($l=1$) mixed modes. We use this example to highlight that we can distinguish between different types of stellar oscillations, e.g., transient (damped, stochastically-excited) and strictly periodic (undamped, heat-driven). We also illustrate the technique of dividing a time-series into chunks to further examine the transient versus periodic nature of stellar oscillations. The harmonic F-test combined with \texttt{mtNUFFT} is implemented in the public Python package \texttt{tapify} (\url{https://github.com/aaryapatil/tapify}), which opens opportunities to perform detailed investigations of periodic signals in time-domain astronomy.
\end{abstract} 

\section{Introduction}
Periodic signals are commonly found in several science and engineering disciplines. Searching for them in time-series data and estimating their frequencies is thus an important statistical problem \citep[][]{percival_1993, bloomfield_2004}. In particular, this problem is ubiquitously found in the field of asteroseismology, the study of the interior of stars using time-series observations of stellar oscillations \citep{aerts_2010_book, chaplin_2013, garcia_2019, aerts_2021}. In this paper, we focus on detecting and characterizing periodic (or almost periodic) signals in asteroseismic time-series data using harmonic analysis.

\begin{figure*}
    \centering
    \includegraphics[width=1\linewidth]{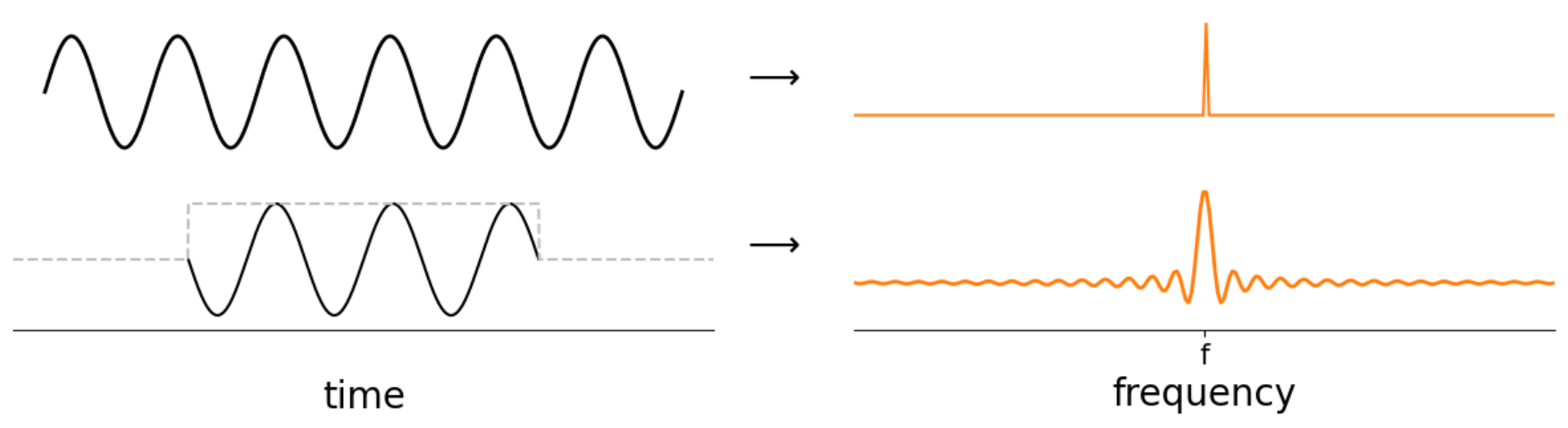}
    \caption{The effect of windowing a sinusoidal signal. The left panels show that using a finite-length time-series is equivalent to multiplying an infinite length time-series with a rectangular window. The right panels show that in the Fourier domain, windowing results in a sinc function instead of a delta function at the frequency of the signal. Essentially, multiplication in time-domain is convolution in Fourier-domain and so the delta function is convolved with the window transform (sinc function).}
    \label{fig:sinc}
\end{figure*}

Harmonic analysis generally refers to the study of periodic signals, which can either be a sinusoidal oscillation at fundamental frequency $f_0$ or those at multiples of $f_0$, i.e., harmonics \citep[][]{walker_1971}. In the Fourier (or frequency) domain, a sinusoid, $A \cos(2 \pi f_0 t + \phi)$ with constant amplitude $A$, frequency $f_0$, and phase $\phi$, that is continuously observed over an infinite time duration is represented by a Dirac delta function at $f_0$. Thus, the Dirac delta function is the Fourier transform or the power spectrum of a sinusoidal signal in time. The statistics literature refers to such strictly periodic or harmonic signals as \textit{line} components\footnote{A ``line" in a power spectrum was named as such because it is similar to e.g., an absorption or emission line in an optical spectrum.}.

When a sinusoidal signal of frequency $f_0$ is continuously-sampled over a finite time duration, some of its power leaks to neighbouring frequencies. In Figure \ref{fig:sinc}, we show that a finite-length sinusoid is a sinc function in Fourier space \citep[refer to][for more details]{vanderplas_2018}. While a sinc function in Fourier space represents an underlying line component (delta function in frequency), distinguishing it from other types of signals (e.g., a Lorentzian in frequency) is challenging. This task becomes even more challenging when the sinusoid is discretely-sampled and the time samples are irregularly spaced. Regular time-sampling results in leakage of power at $f_0$ to distant frequencies whereas irregular sampling results in a ``noisy" or randomized leakage pattern \citep[refer to Figures 7 to 10 in][]{vanderplas_2018}. 

The occurrence of power belonging to $f_0$ at neighbouring or distant frequencies $f + \Delta f$ is referred to as \textit{spectral leakage}. Overall, this phenomenon makes it difficult to detect and characterize line components in time-series, and the field of harmonic analysis focuses on tackling this problem.

In addition to spectral leakage, noise makes it is challenging to detect line components in time-series. Traditional harmonic analysis approaches assume that a time-series is stationary with uniform sampling and Gaussian and/or white noise \citep{percival_1993}. Here, a stationary process refers to that whose statistics are independent of the time origin, and white noise refers to a process whose power spectrum is constant. For example, according to Bayesian probability theory, the Lomb-Scargle \citep[LS;][]{scargle_1982} periodogram is optimally suited to detect a stationary strictly periodic (sinusoidal) signal embedded in Gaussian noise \citep{jaynes_1987}. However, real time-series rarely follow the above assumptions, and in this paper we focus on detection of line components in the presence of colored noise. Colored noise represents that the noise (or the continuum) power spectrum is colored, with varying power across frequencies. Such noise is commonly encountered in times-series observations of solar-type and red giant stars, which makes it difficult to characterize oscillations in these stars \citep{chaplin_2013, garcia_2019}.

The power spectra underlying solar-type and red giant stars comprise of solar-like oscillations called pressure (p-)modes, which are resonant sound waves that are stochastically excited and intrinsically damped in the convective envelopes of the stars \citep{aerts_2010_book, chaplin_2011}. These modes are superimposed on a dominant granulation background, a continuum signal whose power decreases with increasing frequency \citep{kallinger_2014} and is statistically represented as red noise \citep{basu_2020}. In Patil et al. (2024a), we developed the \texttt{mtNUFFT} power spectrum estimator and applied it to efficiently and accurately estimate the Lorentzian shapes of p-modes that sit on top of the granulation background. \texttt{mtNUFFT}, or multitaper Non-Uniform Fast Fourier Transform, estimates the power spectrum of a quasi-regularly sampled time-series using multiple tapers called Discrete Prolate Spheroidal Sequences \citep[DPSS;][]{slepian_1978}. Such multi-tapering provides an estimate with reduced bias and variance as compared to the LS periodogram, which helps improve p-mode parameter inference. In this paper, we extend \texttt{mtNUFFT} to perform harmonic analysis using the multitaper F-test \citep{thomson_1982}, thereby improving line component detection and characterization in time-series with quasi-regular time-sampling. We show that this extension helps distinguish between p-modes and other types of modes, e.g., mixed modes in red giants and gravity (g-)modes in massive stars.

Red giant stars exhibit non-radial mixed modes that behave like acoustic waves in the envelope and gravity waves in the core \citep{arentoft_2008, bedding_2010, chaplin_2010, beck_2011}. Gravity waves or g-modes are low-frequency, long-lifetime oscillations driven by the buoyancy force in the radiative cores of low-mass stars (or envelopes of massive stars) \citep{aerts_2010_book, aerts_2021}. Following main sequence evolution, the contraction of the helium core and the expansion of the hydrogen envelope leads to an increase in buoyancy frequency (upper bound on g-mode frequencies) and decrease in acoustic cut-off frequency (lower bound on p-mode frequencies). This results in the propagation of some p and g-modes to the core and envelope, respectively, which then leads to mode coupling. These mixed modes have a complex structure in frequency space that should be some combination of Lorentzian profiles of p-modes and line components of g-modes. Thus, we use the harmonic F-test to automatically detect mixed/g-modes with (almost) strictly periodic nature and precisely estimate their frequencies. 

Note that we can also use the F-test to detect line components in asteroseismic time-series that are due to processes external to the star \citep[][]{auvergne_2009, borucki_2010, ricker_2015}. For example, we demonstrate that the F-test has the potential to automatically detect harmonics of certain types of exoplanet transits.

As shown in Figure \ref{fig:sinc}, one can detect periodic signals in a time-series by estimating its underlying power spectrum, and then finding signatures in the power spectrum that resemble line components (or sinc functions). Due to this, the terms harmonic analysis and (power) spectrum estimation\footnote{spectrum refers to power spectrum in this paper} are used interchangeably. However, harmonic analysis is a much older field and more specific to the detection of strictly or nearly periodic signals as compared to spectrum estimation, which deals with power spectra that include both line components and continuum, with an emphasis on the continuous part \citep[see][for more details]{thomson_1982}. In Patil et al. (2024a), our focus was on using multi-tapering for spectrum estimation, whereas the goal of this paper is to perform harmonic analysis using the F-test.

This paper is organized as follows. Section~\ref{subsec:mtNUFFT} provides a brief overview of power spectrum estimation using the \texttt{mtNUFFT} (we encourage the reader to see Patil et al 2024a for more details). Section \ref{sec:methods} describes how we extend \texttt{mtNUFFT} to perform harmonic analysis of time-series with quasi-regular sampling. In particular, the extension to \texttt{mtNUFFT} is done using three different approaches. First, we combine the multitaper F-test for regular time-sampling with \texttt{mtNUFFT} in Section~\ref{subsec:f_test}, and detect strictly periodic signals (e.g., exoplanet transit harmonics and mixed modes) in Kepler time-series data \citep{borucki_2010, koch_2010}. Second, we explain how one performs this test multiple times (selective inference) to test the presence of periodic signals at different frequencies (Section~\ref{subsec:mht}). Third, we develop the approach of dividing time-series into chunks of different lengths to understand the temporal behaviour of periodic signals detected using the F-test (Sections \ref{subsec:var_F-test} and \ref{subsec:chunks}). We provide these extensions as part of the \texttt{tapify} package (\url{https://github.com/aaryapatil/tapify}) introduced in Patil et al. (2024a).

\section{\texttt{mtNUFFT} Review}\label{subsec:mtNUFFT}
In Patil et al. (2024a), we developed \texttt{mtNUFFT}, a method that uses multiple Discrete Prolate Spheroidal Sequences (DPSS) to compute a multitaper power spectrum estimate with reduced variance and bias as compared to the LS periodogram. Here, we summarize the steps to compute the \texttt{mtNUFFT} power spectrum estimate of a time-series $\mathbf{x} = \{x_n\}$ with quasi-regular time-sampling $\mathbf{t} = \{t_n \mid n = 0,...,N-1\}$.

\begin{enumerate}
    \item Choose a time-bandwidth product $NW > 1$ that is suitable for a given study. One makes this choice based on a trade-off between variance, spectral leakage, and frequency resolution. As $NW$ increases, the variance and out-of-band spectral leakage\footnote{leakage of power at frequency $f$ outside the interval $(f-W, f+W)$, where $W$ is the bandwidth} (bias) reduces, both improving the statistical stability of the estimator. However, an increase in $NW$ makes the frequency resolution worse (increased local bias\footnote{The power spectrum is averaged over the interval $(f-W, f+W)$}). The bandwidth (or frequency resolution) for a given $NW$ is $NW \times f_\mathcal{R}$ where $f_\mathcal{R}$ is the Rayleigh resolution given by
    \begin{equation}\label{eq:Rayleigh}
    f_\mathcal{R} = \frac{1}{N \Delta t} = \frac{1}{T}
\end{equation}
    Rayleigh resolution is the smallest frequency spacing between two resolved signals. This frequency spacing is higher for the multitaper estimator.
    \item Choose a number, $K$, of orthonormal DPSS tapers $\mathbf{v}_k(N, W)$, where
    \begin{equation}
        K \lessapprox 2 NW.
    \end{equation} In practice, one chooses $K \le 2NW-1$ as too many tapers can lead to badly-biased estimates.
    \item Estimate the single-tapered, independent power spectrum estimates $\hat{S}_k(f) = \left|y_k(f)\right|^2$, where
    \begin{equation}\label{eq:mt_eigen}
         y_k(f) = \sum_{n=0}^{N-1} v_{k, n}^{\star} x_n e^{-i 2 \pi f t_n}
    \end{equation} are the eigencoefficients of the time-series $\mathbf{x}$ and $v_{k, n}^{\star}$ are the regularly-sampled tapers $\mathbf{v}_k(N, W)$ interpolated to times $\mathbf{t}$\footnote{Interpolation is only needed if the time sampling is irregular} \citep{springford_2020}. The sum in the above equation is computed using the adjoint \texttt{NUFFT}, which computes  Fourier series coefficients of $\mathbf{x}$ on a uniform frequency grid. Note that one can also zero-pad $\mathbf{x}$ before applying the adjoint \texttt{NUFFT}, as zero-padding reduces the frequency grid spacing. 
    \item Average the above single-tapered estimates to obtain the multitaper power spectrum estimate 
    \begin{equation}\label{eq:mt_spec}
    \hat{S}^{(\mathrm{mt})}(f) = \frac{1}{K} \sum_{k=0}^{K-1} \left|y_k(f)\right|^2.
\end{equation}
\end{enumerate}

We refer the reader to Patil et al. (2024a) and references therein for more details on using the \texttt{mtNUFFT}, especially in the context of asteroseismology.

In addition to power spectrum estimation, the DPSS allow us to determine whether a strictly periodic, sinusoidal signal (line component) is present at a given frequency. Particularly, a hypothesis test called the variance F-test utilizes these sequences to infer whether the power at a frequency is contributed by a single sinusoidal-shaped periodic signal. This F-test is described in the next section.

\section{Methods}\label{sec:methods}
The following sections describe the new methodology introduced in this paper. To aid with explanations, we apply our methods to the Kepler time-series data \citep{koch_2010, borucki_2010}. In particular, we use the case study of the Kepler-91 red giant, which was also used in Patil et al. (2024a). We thus refer the reader to this paper for more information on the pre-processing of the Kepler-91 time-series data.

\subsection{Harmonic F-test}\label{subsec:f_test}
\begin{figure*}
    \centering
    \includegraphics[width=1\linewidth]{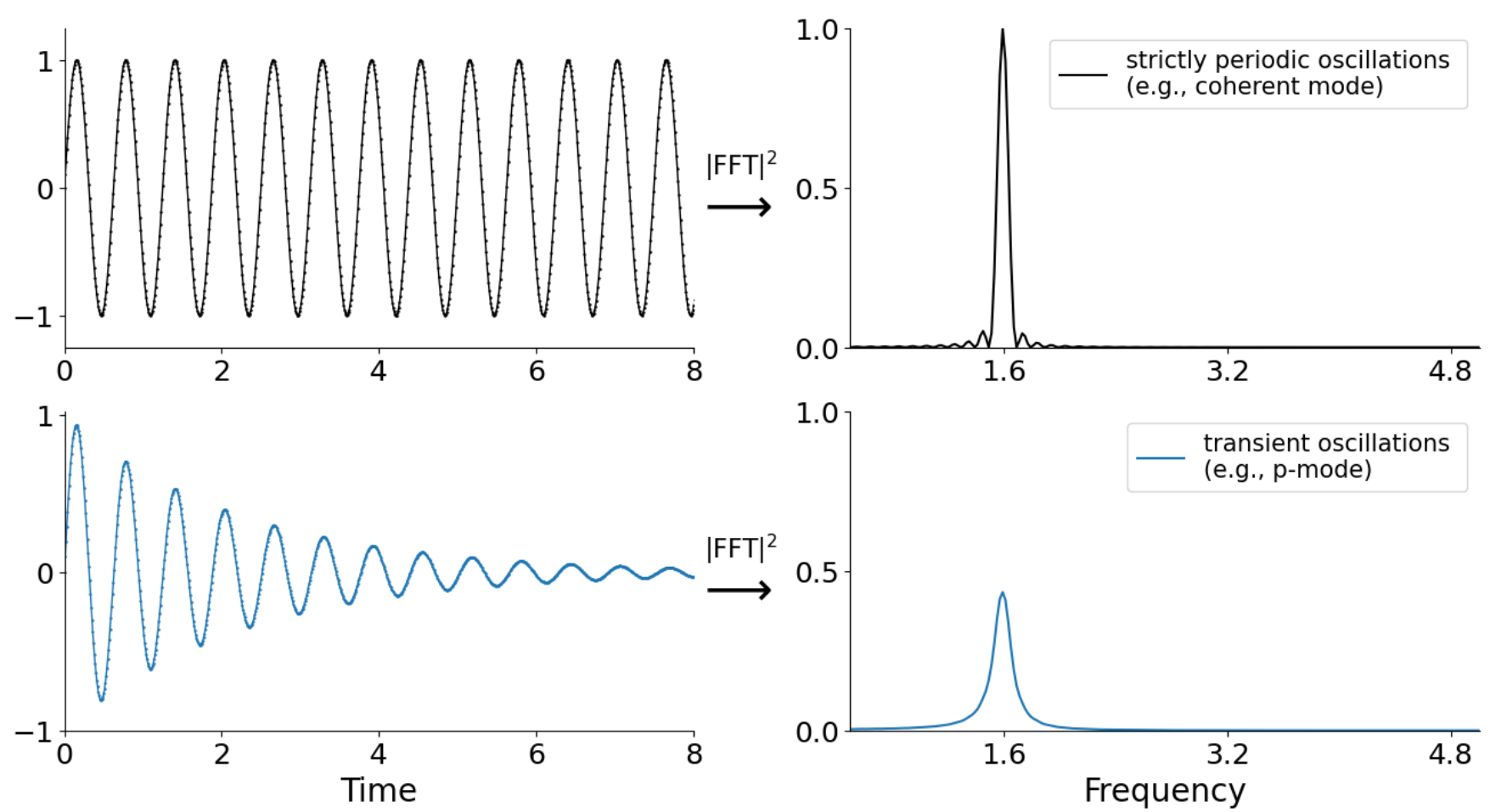}
    \caption{Comparison between sinusoidal, strictly periodic (top panel) and transient oscillations (bottom panel). The left and right panels show time and Fourier domains respectively. In Fourier space, the strictly periodic or harmonic oscillation (e.g., coherent mode) is seen as a peak at the frequency of the oscillation, and the transient damped harmonic oscillations (e.g., p-modes) have Lorentzian frequency peaks with widths representing damping rates.}
    \label{fig:periodicity}
\end{figure*}

In asteroseismology, determining whether a mode is strictly periodic informs us about the mode excitation mechanism. For example, the damping of p-modes makes them transient in nature and adds some quasi-periodicity to their signals.\footnote{In general, quasi-periodic signals have large variation in frequency compared to transient modes; their analysis is out of scope here.}
On the other hand, g- and coherent modes have long and quasi-infinite lifetimes respectively, and therefore resemble strictly periodic signals with sinusoidal shapes. We illustrate these types of oscillations in Figure \ref{fig:periodicity} and their corresponding frequency domain representations using the classical periodogram. 

Strictly periodic, sinusoidal signals are observed as single-frequency peaks in the Fourier domain, which are convolutions of delta functions with the rectangular window function of a time-series (refer to Figure 6 in \citealt{vanderplas_2018}). On the other hand, a transient or damped harmonic oscillation is observed as a broad Lorentzian peak in the frequency domain whose width depends on the damping rate (Figure~\ref{fig:periodicity}). Here damping leads to deviations from strict periodicity. Thus, we can distinguish between different asteroseismic modes in the Fourier domain.

In the case of solar-like oscillators and red giants, our aim is to detect Lorentzian profiles of p-modes on top of a continuous power spectrum composed of stationary noise, granulation and/or magnetic backgrounds. In massive stars and other types of oscillators, we are interested in finding g- or undamped modes that are close to strictly periodic/sinusoidal and show up as peaks in power spectra. We thus develop a harmonic analysis method based on the multitaper F-test \citep{thomson_1982} to precisely detect the frequencies of sinusoidal signals embedded in power spectra, and also distinguish them from damped oscillations. 

\begin{figure*}[t]
    \centering
    \includegraphics[width=\linewidth]{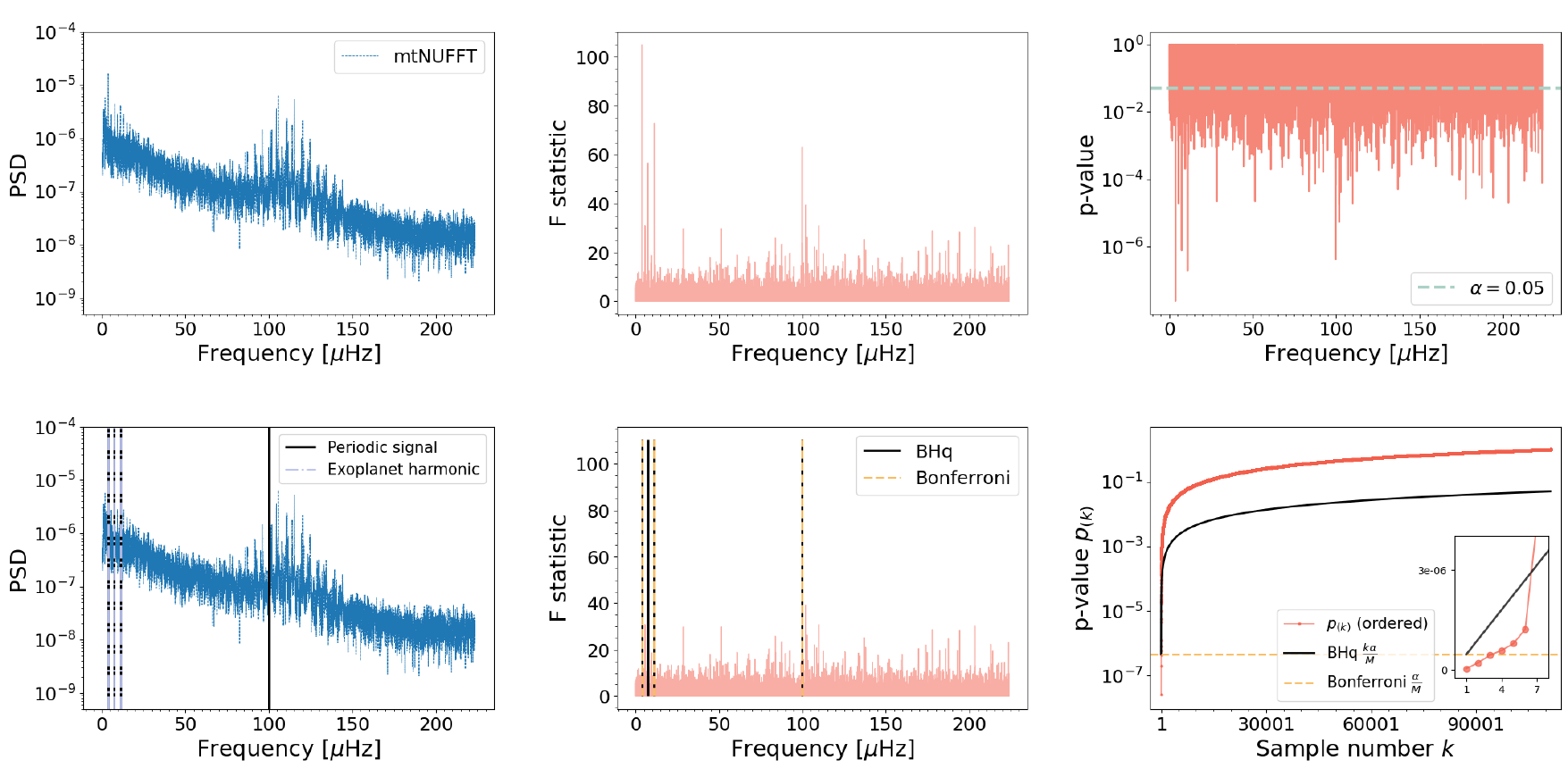}
    \caption{Schematic diagram showing the detection of strictly periodic signals in the Kepler-91 time-series using multitaper F-test. First, we show the \texttt{mtNUFFT} periodogram with $NW=4$ and $K=7$ (top left) and its corresponding F-statistic estimates (top middle). Then, we test the p-values of the F-test for significance (top right shows an example $\alpha = 0.05$ level). Since we are testing multiple hypotheses, we perform selective inference by comparing the sorted p-values with the threshold curves of the Bonferroni and BHq procedures (bottom right). The inset in the bottom right panel zooms into the smallest p-values and shows that BHq rejects more hypotheses than Bonferroni. Finally, we plot the detected periodic signals along with the F-statistic estimates and the \texttt{mtNUFFT} periodogram (bottom middle and left). It is interesting to note that three BHq detected signals coincide with harmonic features that we expect to see due to the known transiting exoplanet Kepler-91b \citep{batalha_2013}.}
    \label{fig:F-test}
\end{figure*}

\subsubsection{F-test for Regularly-sampled Times}
\cite{thomson_1982} develops the analysis-of-variance F-test for evenly-sampled time-series that estimates the significance of a periodic component embedded in coloured noise. It builds on top of the multitaper power spectrum estimate described in Patil et al. 2024a. Essentially, it computes a regression estimate of the power in a periodic signal of frequency $f$ using the eigencoefficients $y_k(f)$ of the regularly-sampled time-series $\mathbf{x}$ (refer to Equation \ref{eq:mt_eigen}) and compares it with the background signal using the following $F$ variance-ratio

\begin{equation}\label{eq:F-test}
    F(f) = \frac{(K-1) \left|\hat{\mu}(f)\right|^2 \sum\limits_{k=0}^{K-1} \left|U_k(N, W; 0)\right|^2}
    {\left|\sum\limits_{k=0}^{K-1} y_k(f) - \hat{\mu}(f) U_k(N, W; 0)\right|^2}.
\end{equation} Here $U_k(N, W; 0)$\footnote{$U_k(N, W)$ is called the discrete prolate spheroidal wave function} is the Discrete Fourier Transform (DFT) of the $k$th order DPSS taper $\mathbf{v}_k(N, W)$ at frequency $f=0$, and $\hat{\mu}(f)$ is the mean estimate of the amplitude of the periodic component at $f$ given by regression methods as

\begin{equation}\label{eq:F-test_regr}
    \hat{\mu}(f) = \frac{\sum\limits_{k=0}^{K-1} U_k(N, W; 0) \, y_k(f)} {\sum\limits_{k=0}^{K-1} {U_k(N, W; 0)}^2}.
\end{equation}

The F-statistic in Equation \eqref{eq:F-test} follows an F-distribution with $2$ and $2K - 2$ degrees of freedom under the null hypothesis that there is no strictly periodic signal at frequency $f$. Here, strictly periodic refers to a signal with stable (or coherent) frequency, phase, and amplitude, i.e., sinusoidal signals.

Notably, the F-test makes use of the phase information in the eigencoefficients $y_k(f)$, which are complex DFTs of DPSS tapered time-series data (refer to Patil et al. 2024a). Particularly, the $y_k(f)$ have complex Gaussian distributions under the F-test null hypothesis. Due to this extra information, the F-test is extremely sensitive to (and preferentially picks) strictly periodic signals that resemble peaks in the Fourier domain. In the context of asteroseismology, such signals generally represent undamped modes or g-modes. On the other hand, the frequencies of transient oscillations shift across a bandwidth surrounding a central frequency, e.g., a stochastically excited p-mode with intrinsic damping is described by a Lorentzian in frequency space (Figure~\ref{fig:periodicity}).

\subsubsection{F-test for Quasi-regularly-sampled Times}
We extend the Thomson F-test to quasi-regular time sampling using the eigencoefficients $y_k(f)$ computed for the \texttt{mtNUFFT} periodogram in Equation \eqref{eq:mt_eigen}. This extension is available in the \texttt{tapify} package as a boolean argument \texttt{Ftest} to the \texttt{MultiTaper.periodogram} function. We refer the reader to the documentation at \url{https://tapify.readthedocs.io/} for more details.

It is important to note that it is necessary to significantly zero pad the adjoint \texttt{NUFFT} that computes the eigencoefficients $y_k(f)$ to ensure that the frequency grid spacing is small enough to detect all present sinusoidal signals. Without zero padding, one runs into the issue of missing a signal that lies between two frequency grid points. We thus zero pad to $M \ge 5N$ (refer to Patil et al. 2024a for details). The exact amount of padding depends on the application, and can be an orders of magnitude higher than $N$. It is good practice to start with a large $M$ to detect all signals, and then refine the Fourier estimate by using a slow FFT in the region of interest.

Using the F-test along with the \texttt{mtNUFFT} periodogram opens avenues for accurately and precisely estimating the frequencies of asteroseismic modes and potentially some extrinsic features in light curves. To demonstrate this, we apply our F-test to the Kepler-91 time-series and show the results in Figure \ref{fig:F-test}, which we discuss in detail in the following section.

\subsection{Multiple Testing Problem}\label{subsec:mht}
Each frequency in the multitaper power spectrum estimate has an associated F-statistic, whose p-value determines the level of significance. If we test all these frequencies individually for significance, we run into the \textit{multiple testing} problem. To understand this, consider the \texttt{mtNUFFT} periodogram in Figure \ref{fig:F-test} which has a total of 111,360 frequencies. For each frequency $f$, we either accept or reject the F-test null hypothesis by testing at the standard 5\% significance level. Let us assume that there are 60 strictly periodic signals amongst the 111,360 frequencies. Even in the best case scenario that our method rejects the null hypothesis for all 60 signals, it is also expected to flag 5\% of the remaining 111,300 quasi- or non-periodic signals as significant, i.e. $0.05*111,300 = 5565$ falsely rejected null hypotheses or false positives \citep{janson_2017}.

To tackle this, we use \textit{selective inference} procedures that rigorously control the number of false positives when testing a (selected) set of hypotheses instead of doing so on a per-hypothesis basis. Selective inference procedures are generally divided into two categories, depending on the selective error rate they control:
\begin{enumerate}
    \item Familywise Error Rate (FWER) is the probability of having at least one false positive.
    \item False Discovery Rate (FDR) is the proportion of false positives among rejected null hypotheses \citep{benjamini_1995}.
\end{enumerate} FWER-controlling procedures limit the presence of even a single false positive, whereas FDR-controlling ones allow for a small number of them. Thus, FWER control is more conservative than FDR. While several procedures are available to control each error rate, we use the Bonferroni and Benjamini Hochberg (BHq) procedures to control FWER and FDR respectively. We define these as follows.

Let $H_1, \dotsc, H_M$ be a set of null hypotheses with corresponding p-values $p_1, p_2, \dotsc, p_M$, where $M$ is the total number of null hypotheses of which some are true nulls and some are false. The false nulls are unknown to us but should be rejected by the FWER-controlling procedures. In this paper, the null hypothesis $H_1$ represents that no periodic signal is present at the frequency $f_1$, and its corresponding p-value $p_1$ tells us the probability of obtaining an F-statistic at least as extreme as $F(f_1)$. Then, the Bonferroni and BHq procedures are given by

\begin{enumerate}
    \item Bonferroni procedure controls the FWER at a significance level $\alpha$ using a fixed threshold. Particularly, Bonferroni rejects null hypotheses with p-values below the threshold
    $$ p_k \leq \frac{\alpha}{M}.$$
    \item Benjamini Hochberg (BHq) controls the FDR at a significance level $\alpha$ using an adaptive threshold. Particularly, given a set of p-values sorted in ascending order $p_{(1)} \leq p_{(2)} \leq \dotsc \leq p_{(M)}$, BHq rejects corresponding hypotheses $H_{(1)}, \dotsc, H_{(\hat{k})}$ where $\hat{k}$ is defined as 
    $$\mathrm{the \; largest} \; k \; \mathrm{for \; which \;} p_{(k)} \leq \frac{k \alpha}{M}.$$
\end{enumerate} Note that instead of controlling the FWER/FDR at the 5\% significance level, one could apply the $1/N$ level suggested by \cite{thomson_1990} as a rule-of-thumb for the F-test.

An F-test detection does not necessarily have to coincide with a clean peak in the multitaper power spectrum estimate at the same frequency. For instance, a strictly periodic signal with frequency $f_0$ and a small amplitude $A$ will have a large associated F-statistic $F(f_0)$ but a small PSD $\hat S^{(\mathrm{mt})}(f_0)$. However, the credibility of an F-test detection increases when it is accompanied by a power spectral peak. The $1/N$ significance level for the Kepler-91 time-series with sample size $N = 44544$ results in $17$ signal detections that do not look significant when compared with the \texttt{mtNUFFT} peaks. We thus prefer the more conservative approach of controlling FWER or FDR in this study, but note that one could use the $1/N$ level for simplicity. 

In Figure \ref{fig:F-test}, we use both the Bonferroni and Benjamini Hochberg (BHq) procedures for controlling the FWER and FDR respectively at the 5\% significance level ($\alpha=0.05$). The inset in the bottom right panel of this figure shows how the two procedures work geometrically. It shows the plot of ordered p-values $p_{(k)}$ versus sample number (or index) $k$ in pink. These p-values are then compared with the threshold curves of the two procedures. The Bonferroni procedure checks $p_{(k)}$ in an ascending order until a p-value goes above its fixed threshold (as shown by the dotted orange line). On the other hand, BHq checks $p_{(k)}$ in a descending order, i.e., from largest to smallest p-values, and finds the first time a p-value goes below the adaptive threshold (as shown by the solid black line). We observe in the figure that BHq rejects six hypotheses whereas Bonferroni rejects three, and decide to choose the BHq discoveries for broader coverage of periodic signal detections. 

Our procedure detects four \textit{potential} strictly periodic signals, which we follow-up to understand their physical nature. Note that we see four BHq signals instead of six due to splittings resulting from zero padding.  The first three of these signals are at frequencies ${\approx}2/6.25, 4/6.25, 6/6.25$ cycles/day, which we expect to see due to the known transiting exoplanet, Kepler-91b, of period $6.246580 \pm 0.000082$ days \citep{batalha_2013}. Thus, the F-test detects the Kepler-91b transit harmonics, and provides period estimates from the three detections. We can also estimate the variance (or uncertainty) of our frequency estimates by jackknifing over tapers (described in detail in the Section \ref{subsec:var_F-test}). For example, we obtain an estimate of $6.24648617$ $6.246486 \pm 0.002052$ days from the first harmonic. The uncertainty we obtain is only an order-of-magnitude higher than the most precise orbital period estimates of the Kepler-91b exoplanet \citep[precise estimation is generally performed in the time-domain;][]{lillo_2014}.

\begin{figure*}
    \centering
    \includegraphics[width=\linewidth]{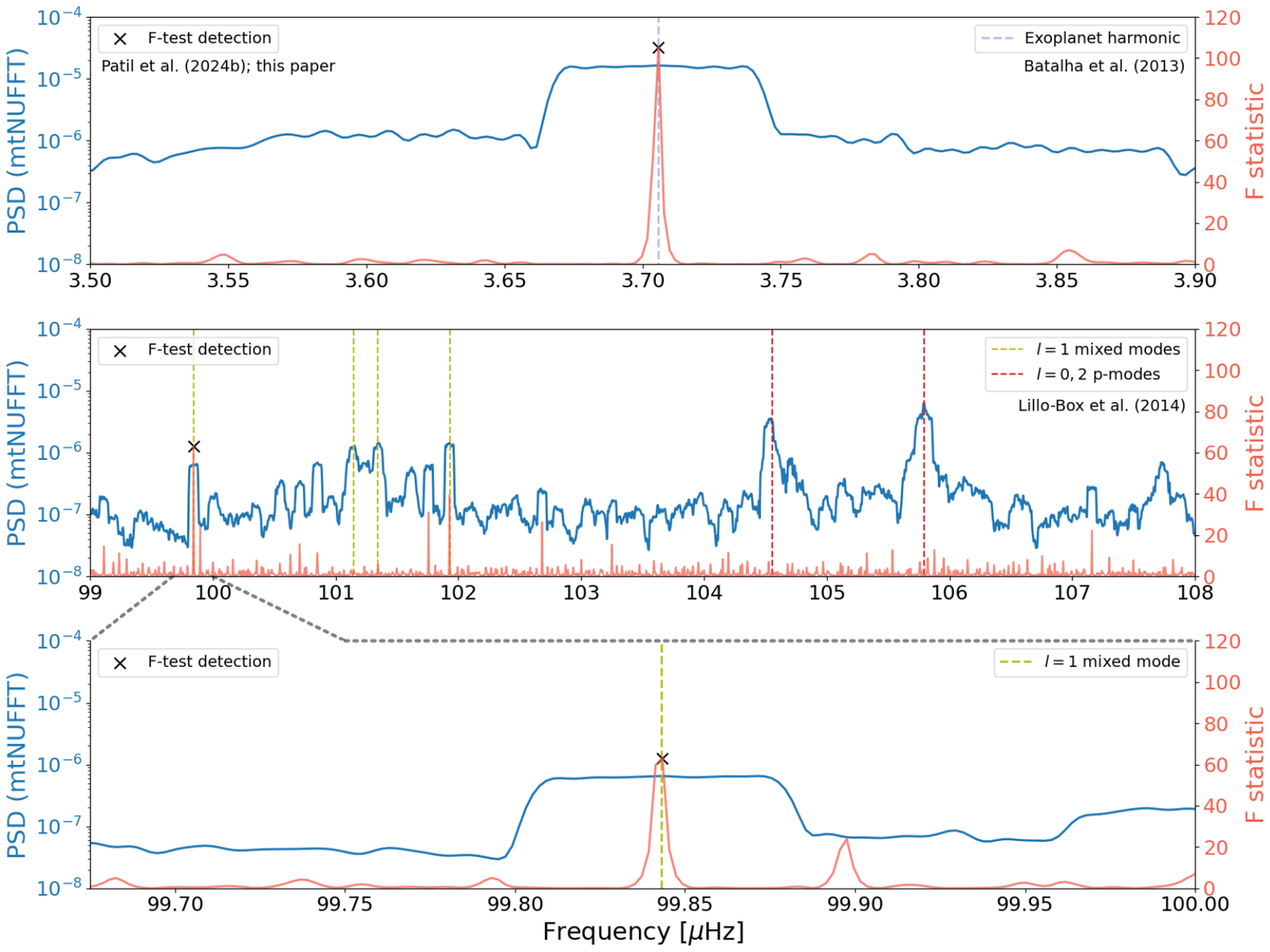}
    \caption{Zoomed-in profiles of the F-test detections that show the rectangular multitaper shape (top and bottom panels) as opposed to the broader Lorentzian (e.g., $l=0, 2$ p-modes in the middle panel). The bottom panel displays a section of the middle panel. \texttt{mtNUFFT} is plotted in blue (left axis is PSD), whereas the F-statistic is plotted in pink (right axis). Two of the F-test detections obtained using the BHq procedure (see Figure \ref{fig:F-test}), coincide exactly with an exoplanet (Kepler-91b) harmonic and an $l=1$ mixed mode. Also, the F-statistic corresponding to the mixed mode at 101.929 $\mu$Hz has a fairly large value. The Kepler-91b period estimate and Kepler-91 mode estimates are from \cite{batalha_2013} and \cite{lillo_2014}, respectively.}
    \label{fig:F-test_detection}
\end{figure*}

The fourth detected sinusoidal signal seems to be situated near a dipole ($l=1$) mixed mode \citep{mosser_2017}. This signal is the only stellar oscillation picked up by the F-test. We can further investigate this detection by zooming in on the profiles of the detected signals as done in Figure \ref{fig:F-test_detection}. The top panel of this figure shows one of the detected Kepler-91b exoplanet harmonics, particularly the one with the largest F-statistic value, and compares it with the harmonic estimate from \citep{batalha_2013}. The middle panel displays the detected $l=1$ mixed mode along with four $l=1$ mode frequencies estimated in \cite{lillo_2014}. This panel also shows the pair of $l=0, 2$ p-modes with the highest PSD. The bottom panel zooms in on the $l=1$ mixed mode shown in the middle panel. We observe that the detected signals in the top and bottom panels have a rectangular multitaper shape as opposed to the broader Lorentzian profile. The rectangular shape represents a delta function with the reduced resolution of the multitaper estimator. Thus, the F-test picks up strictly periodic oscillations (delta functions in Fourier space), which are either exoplanet transit harmonics or mixed mode oscillations in this case study. Why are these types of signals strictly periodic?
\begin{enumerate}
    \item \textit{Exoplanet transits} like those of Kepler-91b are expected to be observed as non-sinusoidal shaped periodic signals in stellar time-series. Such a signal is composed of a fundamental frequency and harmonics, and some of these component harmonics exhibit strictly periodic behaviour.
    \item \textit{Mixed modes} with long lifetimes or dominant g-mode character should resemble line components in stellar power spectra.
\end{enumerate} 

Our method automatically detects Kepler-91b because it has strictly periodic transits that exhibit a square waveform in the time-domain. However, this wave-like transit model generally does not apply to exoplanets; Kepler-91b is a rare exception since it has a transit duration that is long compared with its orbital period. In general, the discovery of exoplanets and precise estimation of planetary properties and orbital parameters is performed by fitting light curves in the time domain. In particular, time domain-based transit timing (variation) methods have provided extremely accurate results over the last two decades \citep[e.g.][]{agol_2005, holman_2007, holman_2010, pozuelos_2023}. However, certain transits such as those of ultra-short period planets have been discovered in the Fourier-domain using transit harmonics in power spectra \citep{sanchis-ojeda_2014}; in these cases, we expect the \texttt{mtNUFFT}/F-test to perform well.

In addition to the mode detection shown in Figure \ref{fig:F-test_detection}, we notice that the mixed mode near $101.929$ $\mu$Hz has a high F-statistic value and would be a significant detection given a smaller BHq significance level $\alpha$. This demonstrates that certain types of mixed modes are detected by the F-test, most likely due to the physical mechanisms underlying these modes, e.g., whether they have more pressure or gravity character. 

The F-test does not detect p-modes, even the high amplitude (PSD) ones shown in Figure \ref{fig:F-test_detection}, since they show up as Lorentzian profiles in Fourier space. The reason behind these profiles is that p-modes are damped (or transient) oscillations rather than strictly periodic; thus, most of the stellar oscillations in Kepler-91 are not detected.

\begin{figure*}[t]
    \centering
    \includegraphics[width=\linewidth]{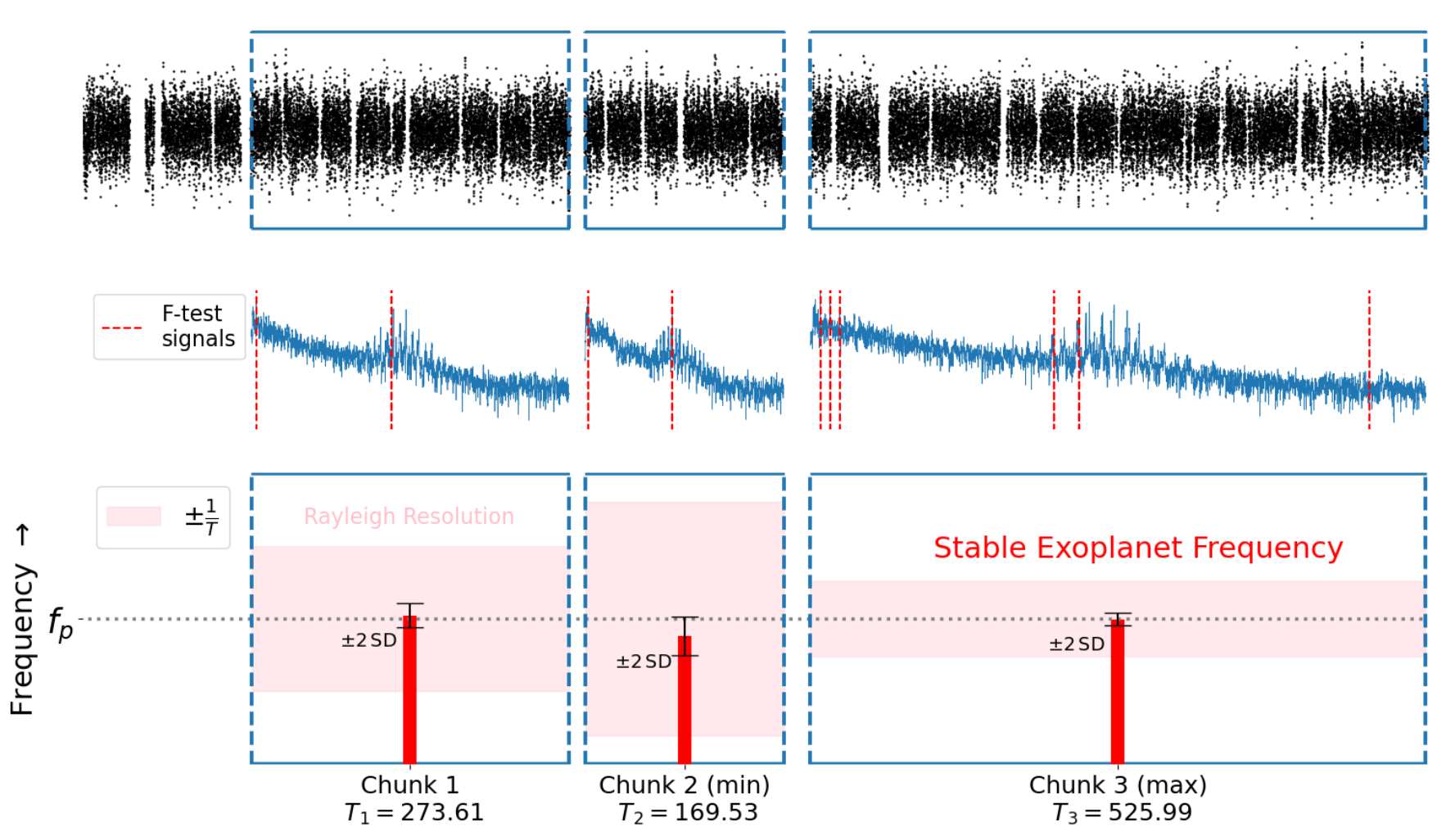}
    \caption{Schematic diagram demonstrating that the F-test frequency estimate $\hat f_p$ corresponding to the Kepler-91b transit harmonic $f_p \approx 2/6.25$ is strictly periodic. The top panel shows how we divide the Kepler-91 time-series into three chunks for studying the time evolution of $\hat f_p$. The middle panel shows the three corresponding \texttt{mtNUFFT} periodograms (blue), and their respective periodic signal detections using the BHq procedure with $\alpha=0.175$ for the F-test (red dashed lines). The bottom panel zooms into the $\hat f_p$ estimates of the three chunks (red bars) with frequency on the y-axis and chunk length ($T$) in days on the x-axis. The estimates are compared with $f_p$ using the two standard deviation jackknife uncertainties (black errorbars) and the Rayleigh resolution $1/T$ (pink) of the chunks. Both comparisons show that the estimates are consistent with $f_p$ and their uncertainties are a fraction of the Rayleigh resolution.}
    \label{fig:F-test_chunk_exo}
\end{figure*}

In the next section, we follow-up our findings by leveraging the frequency precision of the F-test.

\subsection{Precision of the F-test}\label{subsec:var_F-test}
In this section, we demonstrate our follow-up approach of verifying the strict periodicity of F-test detections. We start by assuming an isolated periodic signal (line component) at frequency $f_0$ that is separated from other signals by at least the bandwidth $W$. A good estimate of this frequency would be where the F-test is maximum
\begin{equation}\label{eq:F-test_max}
    \hat{f_0} = \operatorname*{arg\,max}_f F(f)
\end{equation}
As an example, $\hat{f_0}$ corresponds to a harmonic of the Kepler-91b exoplanet transit (Figure~\ref{fig:F-test}).

Under the assumptions of stationary Gaussian locally white noise and moderate \textit{local} SNR of the periodic signal $A \cos(2 \pi f_0 t + \phi)$ with constant amplitude $A$, frequency $f_0$, and phase $\phi$, the variance of the estimate $\hat{f_0}$ is given by
\begin{equation}\label{eq:var_line}
    \mathrm{Var}\{\hat{f_0}\} = \frac{1}{\Xi_K} \frac{6}{{(2 \pi T)}^2} \frac{S_n(f_0)}{S_p(f_0)}.
\end{equation} Above, $T$ is the total time duration of the observed series, and $S_n(f_0)$ and $S_p(f_0)$ are the noise and periodic signal PSD at frequency $f_0$, respectively. Note that $S_p(f_0)$ is equivalent to the periodogram (apparent) PSD of the signal
\begin{equation}\label{eq:signal_line}
    S_p(f_0) = \frac{1}{4} A^2 T.
\end{equation}

$\Xi_K$ in equation~\eqref{eq:var_line} is the variance efficiency of the multitaper power spectrum estimator \citep{thomson_1982}, which is given by
\begin{equation}\label{eq:var_eff}
\Xi_K = \frac{1}{N \sum\limits_{n=0}^{N-1} \left[\frac{1}{K} \sum\limits_{k=0}^{K-1} {[ v_{k, n}(N, W)]}^2 \right]^2}.
\end{equation} We refer the reader to Section \ref{subsec:mtNUFFT} and Patil et al. (2024a) for more information on the above terms.

$\Xi_K$ generally increases as one reduces the variance of the estimate at any frequency as well as the covariance of estimates across frequencies. The inclusion of covariance  is important because one can reduce the variance of a power spectrum estimate by smoothing, but not the covariance. 

Equation \eqref{eq:var_line} is the Cramér-Rao bound \citep[e.g.,][]{rife_1976} with an additional factor of ${\Xi_K}^{-1}$, i.e., it is a few percent larger than the bound \citep{thomson_2007}. Thus, it provides a lower bound on the variance or conversely an upper bound on the precision of the frequency estimate $\hat{f_0}$. We note that for moderately large ${\Xi_K}$ and local SNR $S_p(f_0)/S_n(f_0)$, the standard deviation of $\hat{f_0}$ is a fraction of $1/T$. Take for example the 93.9\% variance efficiency of the multitaper estimator with $NW=3$ and $K=5$. Substituting ${\Xi_K} = 0.939$ in Equation \eqref{eq:var_line} along with $T = 700,000$ years and (local) SNR of 30 results in a standard deviation $\sigma\{ \hat{f_0} \} \approx 10^{-7}$ cycles/year. This standard deviation is approximately one tenth of $1/T$, the Rayleigh resolution. Thus, an important property of the F-test estimator is that it allows us to estimate frequencies of strictly periodic signals with uncertainties smaller than the limiting Rayleigh resolution.

The assumptions underlying Equation \eqref{eq:var_line} are that the periodic signal has constant frequency $f_0$, low enough damping to be mostly within $f_0 \pm W$, and is embedded in Gaussian white noise. However, one almost certainly does not have Gaussian noise, and the local SNR is unknown. Thus, we cannot use the analytical expression for variance in practice. But we can estimate the variance by jackknifing over tapers \citep[see][]{thomson_2007}. There is empirical evidence that the F-test works well for signals isolated by one or two Rayleigh resolutions as opposed to the bandwidth $W$ \citep{thomson_2007}, and the jackknife uncertainties on frequency estimates are expected to be some fraction of Rayleigh resolution as in Equation \eqref{eq:var_line}.

In Patil et al. (2024a), we explain how one can jackknife over tapers to obtain an uncertainty on the multitaper power spectrum estimate $\hat{S}^{(\mathrm{mt})}(f)$, and provide the following equation for jackknife variance

\begin{equation}\label{eq:jk_var}
\widehat{\mathrm{Var}}_J(f) = \frac{(K-1)^2}{K(K - \frac{1}{2})} \sum_{j=0}^{K-1} [\ln \hat{S}^{(\mathrm{mt})}_{\setminus j}(f) - \ln \hat{S}^{(\mathrm{mt})}_{\setminus \bullet}(f)]^2.
\end{equation} for the jackknife variance. Here, the delete-one power spectrum estimates $\hat{S}^{(\mathrm{mt})}_{\setminus j}(f)$ are obtained by omitting the $j$th eigencoefficient $y_j(f)$ from Equation \eqref{eq:mt_spec}, and $\hat{S}^{(\mathrm{mt})}_{\setminus \bullet}(f)$ is estimated by averaging all $\hat{S}^{(\mathrm{mt})}_{\setminus j}(f)$. Similar to this approach, we can jackknife over tapers to estimate the variance of the F-test as follows:

\begin{enumerate}
    \item Estimate $F_{\setminus j}(f)$ by omitting the $j$th eigencoefficient $y_j(f)$ and $U_j(N, W; 0)$ from Equations \eqref{eq:F-test} and \eqref{eq:F-test_regr}.
    \item Estimate $\hat{f}_{0, {\setminus j}}$ for each of the delete-one F-statistic estimates $F_{\setminus j}(f)$ using Equation \eqref{eq:F-test_max}.
    \item Estimate the jackknife variance of the F-test as
    \begin{equation}
        \widehat{\mathrm{Var}}_J\{\hat{f_0}\} = \frac{n-1}{n} \sum_{j=0}^{K-1} \left[\hat{f}_{0, {\setminus j}} -\hat{f}_{0, {\setminus \bullet}}\right]^2
    \end{equation}
\end{enumerate} We used the above procedure to estimate the uncertainty $±0.002052$ days on the period estimate $6.246486$ days of the Kepler-91b exoplanet in Section \ref{subsec:mht}. Here, the quoted uncertainty is the standard deviation, i.e., the square root of $\widehat{\mathrm{Var}}_J\{\hat{f_0}\}$.

\subsection{Dividing Time-series into Chunks}\label{subsec:chunks}

We can further simplify Equation \eqref{eq:var_line} by substituting in Equation \eqref{eq:signal_line}, giving
\begin{equation}\label{eq:var_line_simp}
    \mathrm{Var}\{\hat{f_0}\} \propto \frac{1}{T^3},
\end{equation} which tells us that the variance of the F-test for periodic signals is within a few percent of the Cramér-Rao bound, and so decreases like $1/T^3$. This proportionality demonstrates that reducing $T$ does not significantly increase the variance. Therefore, one can divide the time-series into shorter chunks and apply the F-test to detect periodic signals across these chunks \citep[similar to the Welch method, e.g., in][]{dodson-robinson_2022}. Not only will this reduce the false detection probability (e.g., if you detect a periodic signal in two separate chunks at 99\% significance, you reduce the probability to $10^{-4}$), but also help determine whether a signal is strictly periodic, transient, or a false detection.

\begin{figure*}
    \centering
    \includegraphics[width=\linewidth]{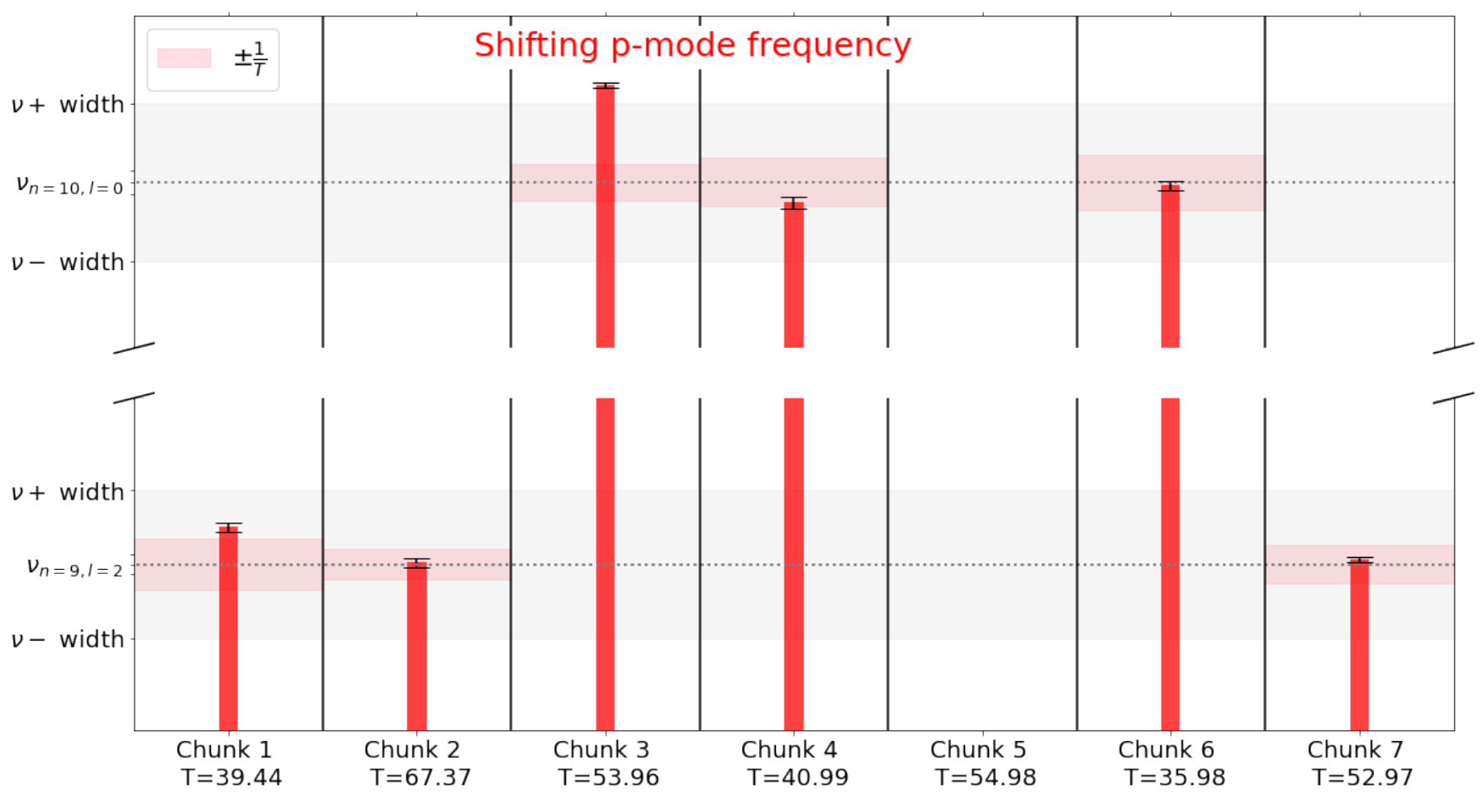}
    \caption{Detection of strictly periodic signals corresponding to two consecutive p-modes, $\nu_{n=9, \, l=2} = 104.557 \; \mu$Hz and $\nu_{n=10, \, l=0} = 105.792 \; \mu$Hz \citep[Table 5 in][]{lillo_2014}.  The red bars indicate the $\hat{\nu}_{n,l}$ detections across seven time chunks, with roughly continuous and even sampling. The x-axis shows the lengths of these chunks ($T$) in days. The y-axis represents frequency, and helps compare the $\hat{\nu}_{n,l}$ estimates (red bars) and their jackknife uncertainties (black errorbars) with the mode frequencies $\nu_{n,l}$ and peak widths \citep{lillo_2014}. The Rayleigh resolution of each chunk is also shown in pink. We see that $\nu_{n=9, \, l=2}$ is detected in chunks 1, 2, and 7, whereas $\nu_{n=10, \, l=0}$ is detected in chunks 3, 4, and 6. The frequency shifts and short-lived detections of modes is reminiscent of damped, short lifetime p-modes.}
    \label{fig:F-test_chunk_p}
\end{figure*}
Solar-like p-mode frequencies vary (or shift) with magnetic activity \citep{baliunas_1995, garcia_2010}, and hence will not be detected (rejected) by the F-test in long time-series. A long time-series would be one that is a significant fraction of the activity cycle period of the star. For example, the solar cycle is approximately 11 years \citep{charbonneau_2010}, and so the F-test does not detect p-modes in a solar time-series that is a few years long (e.g., the Kepler time-series).  Dividing time-series thus allows looking at the nature of stellar oscillations. We describe this as follows:
\begin{enumerate}
    \item A strictly periodic signal will be detected across all time chunks without any significant shifts in its frequency estimates (see Figure \ref{fig:F-test_chunk_exo}, and forthcoming discussion).
    \item Transient p-modes with short lifetimes will undergo frequency shifts \citep{libbrecht_1990} across consecutive chunks. They will also disappear and reappear in detections depending on their lifetimes. To distinguish between the shift of a mode frequency and neighbouring modes, we compare the frequency estimates to named modes and their widths in the literature (see Figure \ref{fig:F-test_chunk_p} and following discussion).
    \item False signals will generally only appear in single isolated time chunks.
\end{enumerate}
Another advantage of dividing time-series into chunks is that we can remove large gaps and analyze continuous quasi-evenly-sampled Kepler observations, thereby controlling spectral leakage and other issues associated with irregular sampling. As Kepler time-series are composed of ${\approx}3$ month quarters, using chunks of ${\sim}90$ days will ensure removal of large gaps. However, with $T = 90$ days, the PSD of the signal $S_p(f_0)$ in question (refer to Equation \ref{eq:signal_line}) reduces significantly and detection becomes difficult. This is especially true for long periods (or low frequencies), as the variance of period estimates goes as
\begin{equation}\label{eq:var_period}
    \mathrm{Var}\{\hat{P_0}\} = \frac{6 P_0^4 S_n(f_0)}{\pi^2 A^2 T^3}
\end{equation}
Therefore, to investigate the low-frequency $\hat{f_0}$ estimate in Figure \ref{fig:F-test}, which corresponds to the Kepler-91b transit harmonic $f_p \approx 2/6.25$ days, we remove large gaps and divide the Kepler-91 time-series into three chunks of lengths $T = 273.61, 169.53, 525.99$ days. We show this in the top panel of Figure \ref{fig:F-test_chunk_exo}. For each of the three chunks, we compute the \texttt{mtNUFFT} periodogram, apply the F-test to detect periodic signals, and control the FDR using the BHq procedure with significance level $\alpha=0.175$, as described in Section \ref{subsec:mht}. The middle panel of Figure \ref{fig:F-test_chunk_exo} shows the three periodograms and their respective signal detections. Note that we use a less conservative significance level (higher $\alpha$ value) for these detections compared to that for the entire time-series because the SNR of such periodic signals is proportional to $T$ (refer to Equation \ref{eq:signal_line}). We then focus on the detection $\hat f_p$ within the range $f_p \pm 2/T_\mathrm{chunk}$; we choose this range because the separability of signal detections by the F-test is on the order of one or two Rayleigh resolutions (as described earlier in this section). Finally, we estimate the variance of $\hat f_p$ by jackknifing over tapers. 

The $\hat f_p$ estimates for the three chunks and their two-standard deviation jackknife uncertainties (${\approx}95$\% confidence interval) are in the bottom panel of Figure \ref{fig:F-test_chunk_exo}. This panel shows that the $\hat{f_0}$ estimate is very stable compared to the Rayleigh resolution as well as the jackknife uncertainties, which we expect from a strictly periodic exoplanet signature. The jackknife uncertainties are ${\approx}1/6$ of the Rayleigh resolution, i.e., they are smaller for longer time chunks.

We also examine the behaviour of the high-frequency p-modes in Figure \ref{fig:F-test_chunk_p} by dividing the Kepler-91 time-series into seven chunks of length ${\sim}60$ days. Using the same method as in Figure \ref{fig:F-test_chunk_exo}, we compute the mtNUFFT periodograms and detect periodic signals using the F-statistic and multi-hypothesis testing. The only difference is that we use a higher $\alpha$ value due to the significant reduction in $T$. We then focus on two consecutive p-modes, $\nu_{n=9, l=2} = 104.557 \; \mu$Hz and $\nu_{n=10, l=0} = 105.792 \; \mu$Hz \citep{lillo_2014} by analyzing corresponding detections. This correspondence is determined through comparison with the mode frequency and its peak width. Across chunks, we see that the detected mode frequencies undergo shifts beyond jackknife uncertainties and the limiting Rayleigh resolution, thereby suggesting the presence of p-modes. In some chunks, one of the modes is not detected at all, but it reappears at a later time; this property might have relations with the lifetime of p-modes. We can thus conclude that dividing time-series into short-length chunks and applying the F-test to consecutive chunks is a powerful tool to detect and characterize asteroseismic oscillations, thereby allowing determination of excitation mechanisms.

The above example shows that the division of the Kepler-91 time-series into chunks of $\sim$60 days length allows detailed analysis of the temporal behavior of p-modes using the F-test. However, the question of how to best choose the length of these chunks remains. We note that it is generally good to use several \textit{short} chunks that are long enough to resolve the relevant modes (refer to Equation \ref{eq:Rayleigh}), but short enough that frequency shift with stellar activity \citep[e.g.,][]{regulo_2016} does not affect the F-test detection. As shown in Equation \eqref{eq:var_line_simp}, the variance of the F-test frequency estimate $\hat{f}_0$ is proportional to $1/T^3$. Therefore, even the smallest frequency shifts will make it difficult to detect a truly periodic signal of frequency $f_0$ in a long time-series (large $T$). Instead, one can use coincidence detection across chunks to choose frequencies. For example, three detections above 90\% give $\sim$99.9\% significance (this percentage varies depending on the number of chunks). Using short-length chunks also helps resolve the problem that a non-Gaussian process (noise), e.g., a Laplace distribution, has effects on a time-series that are randomly placed in frequency.

\section{Results \& Discussion}
In this paper, we extend the \texttt{mtNUFFT} periodogram developed in Patil et al. (2024a) to the multitaper F-test and perform harmonic analysis of asteroseismic time-series. We apply this F-test to the Kepler-91 red giant and show that it allows automatic detection and period estimation of the Kepler-91b exoplanet. This detection is because the F-test is designed to find strictly periodic signals in time-series data such as an exoplanet transit and its harmonics\footnote{the strict periodicity of exoplanet transits is true to first order.}.

Our results suggest that we can use multitapering to further constrain the low-frequency power excess that can help deduce stellar granulation (surface convection), rotational modulation, and other stellar activity \citep[refer to][for a review]{garcia_2019}. Note that some methods \citep[e.g.,][]{kallinger_2014} perform gap filling using interpolation to reduce leakage of the low-frequency granulation signal to high frequencies, but this method itself can lead to some spectral leakage and bias in power spectrum estimates \citep{lepage_2009, springford_2020}. We can instead use \texttt{mtNUFFT} with the F-test to control spectral leakage and precisely estimate granulation backgrounds and rotation peaks/harmonics.

In addition to solar-like oscillators, we can use multitapering to analyze different classes of pulsating stars that span the Hertzsprung-Russell diagram \citep{aerts_2021}. Precise estimation of mode frequencies and lifetimes, whether they are p, g, mixed, or heat-driven undamped modes, opens avenues for detailed studies of stellar interior. We believe that the \texttt{mtNUFFT}/F-test combination could be an improvement over the iterative \textit{prewhitening} \citep{breger_1993} method to search for g or undamped modes in different pulsators \citep[e.g.,][]{reeth_2015a, reeth_2015b, li_2020, aerts_2021}. We explore the detection of g-modes in a forthcoming paper.

In this paper, we also discuss dividing the time-series into chunks and applying \texttt{mtNUFFT} and the F-test to each chunk. We use this approach to verify the truly periodic nature of F-test detections, i.e., to improve harmonic analysis. The Welch periodogram \citep{welch_1967}, on the other hand, follows a similar chunk-based approach, but to obtain a better power spectrum estimate than the classical periodogram. After sectioning the time-series into overlapping chunks, \cite{welch_1967} tapers the chunks, calculates their periodograms, and then averages them to obtain the Welch periodogram estimate \citep[refer to applications in][]{dodson-robinson_2022}.

Power spectrum estimators like the classical, Welch, and LS periodograms are generally well established for stationary processes, but real data are often not strictly stationary \citep{thomson_1982, nason_2006}. These estimators are reasonably robust to non-stationarities, i.e., they can detect a periodic signal with time-varying amplitude and frequency, but their accuracy can be improved by explicitly taking stationarity and non-linearity into account \citep{rahim_2014}. Therefore, multitaper spectral analysis has been extended to include non-stationary processes \cite[e.g.,][]{thomson_1982, hammond_1996}. Our \texttt{mtNUFFT} plus chunk/F-test approach is one such extension that allows detailed analyses of transient (non-stationary) signals such as p-modes.

Essentially, our chunk/F-test approach is similar to a (windowed) short-time Fourier transform \citep[STFT;][]{allen_1977} that decomposes individual chunks into an orthogonal basis of sines and cosines to pick out signals localized in time. However, similar to STFT, there is a trade-off between frequency and time resolution, i.e., the longer the length of the individual chunks, the higher frequency resolution we obtain, but this results in lower temporal resolution. Instead, one can use the wavelet transform \citep{strang_1994} that decomposes a signal using an orthogonal basis of wavelets to improve the overall time and frequency (multi-)resolution. While a detailed comparison between the chunk/F-test and wavelet approach is out of scope of this paper, we note that several studies show that the wavelet approach performs better for truly transient signals but its performance is heavily dependent on the choice of wavelets \citep[][and references therein]{addison_2005}. For asteroseismic analyses of p-modes, wavelets might not be necessary since the power spectrum is sufficient for estimating precise frequencies. More recent approaches in the literature using chunks/windows also test whether a signal is strictly periodic. For example, \cite{hara_2022} use box-shaped time windows to distinguish between planets and stellar activity in radial velocity data.

 In the future, we could further improve our multitaper approach to deal with non-stationarity (refer to the next section for more details).

\section{Conclusion}

We demonstrate that combining \texttt{mtNUFFT} with the multitaper F-test allows us to detect strictly periodic signals in asteroseismic time-series. In the case of the Kepler-91 red giant, the F-test detects mixed modes, stellar oscillations with pressure and gravity character, and the Kepler-91b exoplanet transits, a feature extrinsic to the star. Exoplanet transits are generally detected in the time-domain, but our approach shows that certain types of transits can be directly detected in the frequency domain. More generally, our F-test approach shows promise for frequency estimation of g or coherent modes in intermediate-mass and massive stars. The 
time-domain method currently adopted for these modes \citep[\textit{prewhitening} done in][]{breger_1993} could be replaced by the \texttt{mtNUFFT}/F-test for computational efficiency as well as frequency precision.

In addition, the precision of the F-test allows us to diagnose the periodic versus transient nature of oscillatory signals. For example, we can divide a Kepler time-series into shorter, more continuous chunks and test whether the frequency of a signal remains stable over the duration of the light curve. This technique has prospects for detecting and characterizing different types of asteroseismic modes, thereby providing deeper insights into stellar structure and evolution. 

\section*{acknowledgements}
AAP is supported by an LSST-DA Catalyst Fellowship; this publication was thus made possible through the support of Grant 62192 from the John Templeton Foundation to LSST-DA. This project was supported by the Data Sciences Institute at the University of Toronto (UofT). GME acknowledges funding from NSERC through Discovery Grant RGPIN-2020-04554 and from UofT through the Connaught New Researcher Award, both of which supported this research.

The authors thank the anonymous referee for their thorough and thoughtful feedback. They also thank Conny Aerts for helping design the project, Jo Bovy for providing insightful feedback on the manuscript, and Eric Agol for helpful discussions on the prospects of multitapering for exoplanet detection.

This paper includes data collected by the Kepler mission and obtained from the MAST data archive at the Space Telescope Science Institute (STScI). The data can be found in MAST: \dataset[10.17909/b90b-a558]{http://dx.doi.org/10.17909/b90b-a558}. Funding for the Kepler mission is provided by the NASA Science Mission Directorate. STScI is operated by the Association of Universities for Research in Astronomy, Inc., under NASA contract NAS 5–26555.

\software{\texttt{astropy} \citep{astropy:2013, astropy:2018, astropy:2022}, \texttt{FINUFFT} \citep{barnett_2019}, lightkurve \citep{lightkurve_2018} \texttt{matplotlib} \citep{matplotlib:2007}, \texttt{nfft} \citep{nfft_2017}, \texttt{numpy} \citep{numpy:2020}, \texttt{pbjam} \citep{nielson_2021}, \texttt{scipy} \citep{scipy:2020}.}

\end{document}